\documentclass[aps,prb,twocolumn,superscriptaddress]{revtex4-2}
\usepackage[colorlinks=true,citecolor=blue,linkcolor=blue,urlcolor=blue,]{hyperref}
\usepackage{graphicx, nicefrac, textcomp}
\usepackage[normalem]{ulem}
\usepackage{changes}
\usepackage{chngcntr}
\usepackage{amssymb}
\usepackage{soul}
\usepackage{amsmath}
\newcommand{\RK}[1]{\textcolor{black}{#1}}

\begin{document}
\preprint{APS/123-QED}

\title{SrCu(OH)$_3$Cl, an ideal isolated equilateral triangle spin $S$ = 1/2 model system\\}

\author{Sudip Pal}
\email{sudip.pal111@gmail.com} 
\affiliation{1. Physikalisches Institut, Universität Stuttgart, Pfaffenwaldring 57, 70569 Stuttgart Germany}
\author{Petr Doležal}
\affiliation{Department of Condensed Matter Physics, Faculty of Mathematics and Physics, Charles University, Ke Karlovu 5, 121 16 Prague 2, Czech Republic}  
\affiliation{Institute of Solid State Physics, TU Wien, 1040 Vienna, Austria}
\author{Scott A. Str{\o}m}
\affiliation{Institute of Solid State Physics, TU Wien, 1040 Vienna, Austria}
\author{Sylvain Bertaina}
\affiliation{CNRS, Aix-Marseille Université, IM2NP (UMR 7344), Institut Matériaux Microélectronique et Nanosciences de Provence, Marseille, France}
\author{Andrej Pustogow}
\affiliation{Institute of Solid State Physics, TU Wien, 1040 Vienna, Austria}
\author{Reinhard K. Kremer} 
\email{rekre@fkf.mpg.de}
\affiliation{Max-Planck-Institute for Solid State Research, Heisenbergstraße 1, 70569 Stuttgart, Germany}
\author{Martin Dressel}
\email{martin.dressel@pi1.physik.uni-stuttgart.de} 
\affiliation{1. Physikalisches Institut, Universität Stuttgart, Pfaffenwaldring 57, 70569 Stuttgart Germany}
\author{Pascal Puphal}
\email{puphal@fkf.mpg.de} 
\affiliation{Max-Planck-Institute for Solid State Research, Heisenbergstraße 1, 70569 Stuttgart, Germany}

\date{\today}

\begin{abstract}
We have investigated the magnetic ground state properties of the quantum spin trimer compound strontium hydroxy copper chloride SrCu(OH)$_3$Cl using bulk magnetization, specific heat measurements, nuclear magnetic resonance (NMR),  and electron spin resonance (ESR) spectroscopy. SrCu(OH)$_3$Cl consists of layers with isolated Cu$^{2+}$ triangles and hence provides an opportunity to understand the magnetic ground state of an isolated system of \textit{S} = 1/2 arranged on an equilateral triangle. Although magnetization measurements do not exhibit a phase transition to a long-range ordered state down to \textit{T} = 2 K, they reveal the characteristic behavior of isolated trimers with an exchange of $J = 154$~K. The Curie-Weiss behavior changes 
 around 50--80~K,  as is also seen in the NMR spin-lattice relaxation rate. In zero magnetic field,  our specific heat data establish a second-order phase transition to an antiferromagnetic ground state  below \textit{T}= 1.2 K.  We have drawn a magnetic field-temperature ($H$-$T$) phase diagram based on the specific heat measurements. The ESR data show divergence of the linewidth at lower temperatures, which precedes the phase transition to an antiferromagnetic long-range ordered state with unconventional critical exponents. The temperature variation of the $g$-factor further confirms the antiferromagnetic phase transition and reflects the underlying magneto-crystalline anisotropy of the compound.  
\end{abstract}
                             
\maketitle

\section{\label{sec:level}Introduction:}
Quantum fluctuations arising due to geometrical frustration are currently a vividly studied topic in condensed matter physics. In this respect antiferromagnetically coupled spin \textit{S} =  1/2  entities arranged on a two-dimensional (2D) triangular lattice provide the simplest model system, where \RK{competing} spin-spin exchange pathways may give rise to exotic magnetic ground states. It is now widely accepted, that in zero magnetic field its ground state is a three-sub-lattice 120$^\circ$  order {\color{blue}\cite{Anderson1973,Fazekas1974, Huse1988,Singh1992,Sachdev1992,Szasz2020,Jolicoeur1989, Susuki2013}}. Substantial theoretical effort to study the antiferromagnetic Heisenberg model with additional terms, such as second-neighbor interactions, ring exchanges etc. are found to give rise to other interesting phases including the Dirac quantum spin liquid {\color{blue}\cite{Hu2019}}. Nonetheless, despite the relentless efforts of many decades, our understanding of the ground state of a \textit{S} = 1/2 triangular lattice system is yet incomplete. 

To this end, an investigation of the magnetic properties of isolated antiferromagnetically coupled trimers of spins arranged in equilateral triangles, i.e. the elementary building block of a frustrated system in 2D should be enlightening. The Hamiltonian for an equilateral trimer spin system is given by ($\alpha$ = 1), 
\begin{equation}
H = J[\bf{\alpha\,(S_1\cdot S_2 + S_2\cdot S_3) + S_1\cdot S_3}],
\label{Hamiltonian}
\end{equation}
where $J$ is the antiferromagnetic exchange interaction between three nearest neighbour Heisenberg spins $\textbf{S}_i (i=1,2,3)$ (see also Fig. \ref{Isosceles}){\color{blue}\cite{Kahn1997}}.   $\alpha \neq 1$ describes an  isosceles triangle.

The ground state of a half-integer spin triangle is non-trivial and frustrated. In the case of $S$ = 1/2 on an equilateral triangle, the lowest energy state is the four-fold degenerate $S_{total}$ = 1/2. The ground state is separated from the next excited state of $S_{total}$ = 3/2 with an energy gap of 3$J$/2. Quantum spin trimers exhibit varying quantum phases and non-trivial magnetic excitations {\color{blue}\cite{Waldmann2003,Cao2020,Bera2022}}.

Trimers might be an important ingredient for various other 2D lattices with subtle distortion, as proposed for the kagome \cite{Harris2013}, which was discussed in experiments for Cu$_3$V$_2$O$_7$(OH)$_2\cdot$2H$_2$O \cite{Janson2016}, but doubts \RK{remained} as a 1D chain might be better able to describe the experimental picture \cite{Whangbo2022}. Recent structural investigations of low temperature diffraction on various kagome systems show that a structural instability induces distortions in barlowite, claringbullite~\cite{Henderson2019}, volborthite~\cite{Ishikawa2015} and vesignieite~\cite{Boldrin2016} making it a typical structural motif at low temperatures, in addition to systems that are already distorted at room temperature such as Y-Kapellasite~\cite{Puphal2017}, where theoretical studies show a unique sublattice realization consisting of triangle building blocks ~\cite{Hering2022}. 
Thus understanding the building blocks of such lattices is essential. Moving to real trimer sublattice systems, realization in inorganic compounds is rare and is usually found as linear trimers as found in 9R-type perovskites like Ba$_5$Ru$_3$O$_{12}$ \cite{Basu2020} and its family Ba$_4M$Ru$_3$O$_{12}$ \cite{Nguyen2020} or Ba$_4$Ir$_3$O$_{10}$ \cite{Cao2020}, etc., as well as linear chains that trimerize  as in Na$_2$Cu$_3$Ge$_4$O$_{12}$\cite{Bera2022}.

So far, the best established real isolated $S$ = 1/2 trimer system, La$_4$Cu$_3$MnO$_{12}$, has been studied using bulk magnetization, specific heat at zero magnetic field, and neutron scattering experiments {\color{blue}\cite{Azuma2000,Qiu2005,Wessel2001,Weber2022}}. La$_4$Cu$_3$MnO$_{12}$ undergoes antiferromagnetic order below \textit{T} = 2.6 K, and exhibits a plateau in the magnetization above a critical field of approximately 20 T. However, this compound suffers from disorder that arises from a split Cu site. The scarcity of such compounds, particularly good quality single crystals restricted the experimental realization of further investigations. 

\begin{figure}[h]
\centering
\includegraphics[width=1.0\columnwidth]{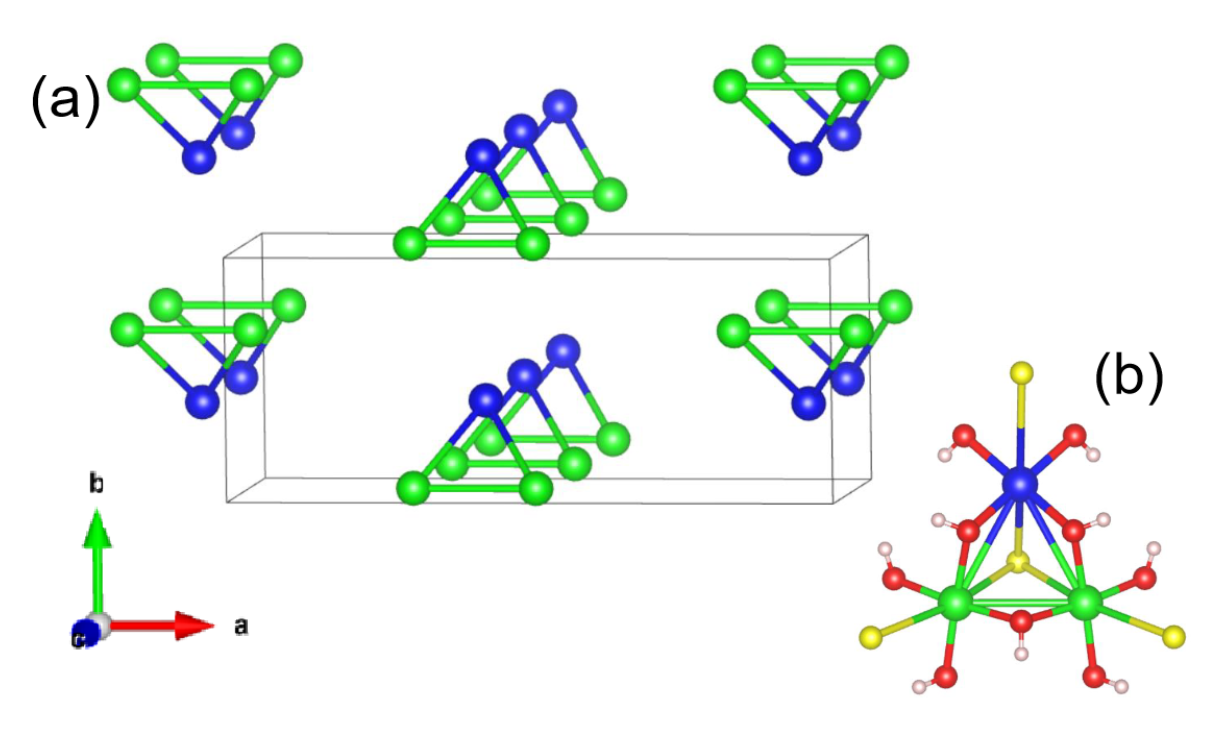}
\caption{Crystal structure of SrCu(OH)$_3$Cl. (a) Arrangement of the Cu isosceles triangles. A unit cell is outlined. (b) Coordination \RK{of Cu} by oxygen and chlorine  atoms. The blue/green, red, yellow spheres represent Cu, O, and Cl atoms, respectively.} \label{Newxrd}
\end{figure}

Here, we report on the magnetic and magneto-thermal properties of the system SrCu(OH)$_3$Cl which features antiferromagnetically coupled  spin \textit{S} = 1/2 trimers arranged in magnetically isolated isosceles triangles.
SrCu(OH)$_3$Cl has firstly been synthesized and its crystal structure has been determined by Zhu \textit{et al.} {\color{blue}\cite{Zhu2014}}. SrCu(OH)$_3$Cl crystallizes in the orthorhombic space group \RK{no.} 31 containing two different Cu atoms. Of interest for its magnetic properties are three Cu$^{2+}$ cations arranged in isosceles triangles which are well separated from each other as displayed in Fig. \ref{Newxrd}. The  Cu$^{2+}$ cations each are coordinated by four oxygen atoms at distances of $\sim$ 2 \AA~ forming equatorial planes containing the Cu $d_{x^2 - y^2}$ orbitals.  Cl atoms are at apical positions at distances of $\sim$ 3.1 \AA. The Cu -- O -- Cu bonding angles amount to  119.4$^{\rm o}$ (2$\times$ and 112.6 $^{\rm o}$ (1$\times$), respectively.
The crystal structure of SrCu(OH)$_3$Cl has been described as consisting of cuprate layers in the {011} planes with the alternating orientation of the triangular planes, thus being different from the parallel arrays in the kagome layers found in the related copper hydroxy chloride family $A$Cu$_3$(OH)$_6$Cl$_2$ \cite{Puphal2018}. First magnetic characterization by Zhu \textit{et al.} indicated rather sizeable exchange interaction between the Cu spins. At high temperatures, the inverse magnetic susceptibility followed a Curie-Weiss behavior with a Curie-Weiss temperature of around -135 K, indicating predominant antiferromagnetic exchange between the Cu moments. Magnetization measurements at 2 K proved saturation with a saturated moment of $\sim$ 1 $\mu_{\rm B}$ indicating a spin $S$ =1/2 ground state with a $g$-factor close to 2.
SrCu(OH)$_3$Cl bears some similarity  to  trimer molecular magnets, where isolated magnetic clusters are either a part of a two- or three-dimensional lattice or are surrounded by non-magnetic ligands {\color{blue}\cite{Dagotto1996,RojasDotti2018}}.

We have grown high-quality single crystals of SrCu(OH)$_3$Cl and investigated the magnetic and thermodynamic properties using bulk magnetization, nuclear magnetic resonance (NMR), electron spin resonance (ESR) spectroscopy and specific heat measurements. The temperature-dependent magnetization, $M(T)$ does not indicate any phase transition down to $T$ = 2 K as reported earlier. However, our specific heat measurements  in zero magnetic field reveals a $\lambda$-type anomaly at  $T \approx$ 1.2 K involving the removal of a magnetic entropy corresponding to a spin $S$ = 1/2 entity.  The temperature dependence of the ESR linewidth and of the $g$-factor supports this finding.  A magnetic phase diagram is constructed from the magnetic field dependence of the specific heat at low temperatures.

\section{\label{sec:level}Experimental details:}
Transparent blue single crystals (see Fig. \ref{xrd}(b)) of SrCu(OH)$_3$Cl were grown \RK{hydrothermally} using a hydroflux method which was modified as compared to the description given in  Ref. \cite{Zhu2014}. A mixture of 5.34 g SrCl$_2$-6H$_2$O, 2.666 g CuCl$_2$-2H$_2$O, 2.4 g LiOH was poured sequentially into a 30 \>RK{ml} Teflon lined autoclave without any further addition of water or stirring and heated with 0.1 K/min to 240$^\circ$C, subsequently held and cooled with an oscillatory profile at a rate of 0.1 K/min. The crystals were finally washed with distilled water. The structural properties of the obtained crystals were characterized using powder x-ray diffraction (PXRD) performed at room temperature using a Rigaku Miniflex diffractometer with Bragg Brentano geometry, Cu K$_\alpha$ radiation, and a Ni filter. Rietveld profile refinements were conducted with the FullProf software suite {\color{blue}\cite{Fullprof}}.
Single crystal diffraction was performed at room temperature using a Rigaku XtaLAB mini II with Mo K$_\alpha$ radiation. The data were analyzed with the CrysAlis(Pro) software and the final refinement done using Olex2 with SHELX.
The magnetization measurements have been performed at temperatures $2\le T \le 300$ K using a superconducting quantum interference device (MPMS3, Quantum Design). 
The specific-heat measurements were carried out, down to $T$ = 400~mK in a \RK{Physical Properties Measurement System} (PPMS, Quantum Design) equipped with the He3 option. We note here that due to very small mass of the individual crystals, the specific heat measurements have been performed on an assembly of a few single crystals \RK{oriented} with magnetic field aligned along the $c$-axis.
The $^1$H NMR experiments were performed in a 14~T PPMS (Quantum Design) cryostat at temperatures 3 – 110~K in an external magnetic field 1.79~T. The spin-lattice relaxation rate $T_1^{-1}$ was determined from stretched exponential fits of the recovery curves.
The ESR spectra were collected in a Bruker X-band spectrometer ($\nu$ = 9.47 GHz) equipped with a continuous He gas flow cryostat working in the temperature range  down to $T$ = 1.8~K. The signal-to-noise ratio of the spectra is improved by recording the field derivative of the absorption spectra ($dP/dH$) using the lock-in technique with field modulation at 100 kHz. The samples were fixed in a quartz tube by paraffin and could be rotated using an automated goniometer.

\section{\label{sec:level}Results and discussions}
\subsection{\label{sec:level}x-ray diffraction}
\begin{figure}[h]
\centering
\includegraphics[width=1.0\columnwidth]{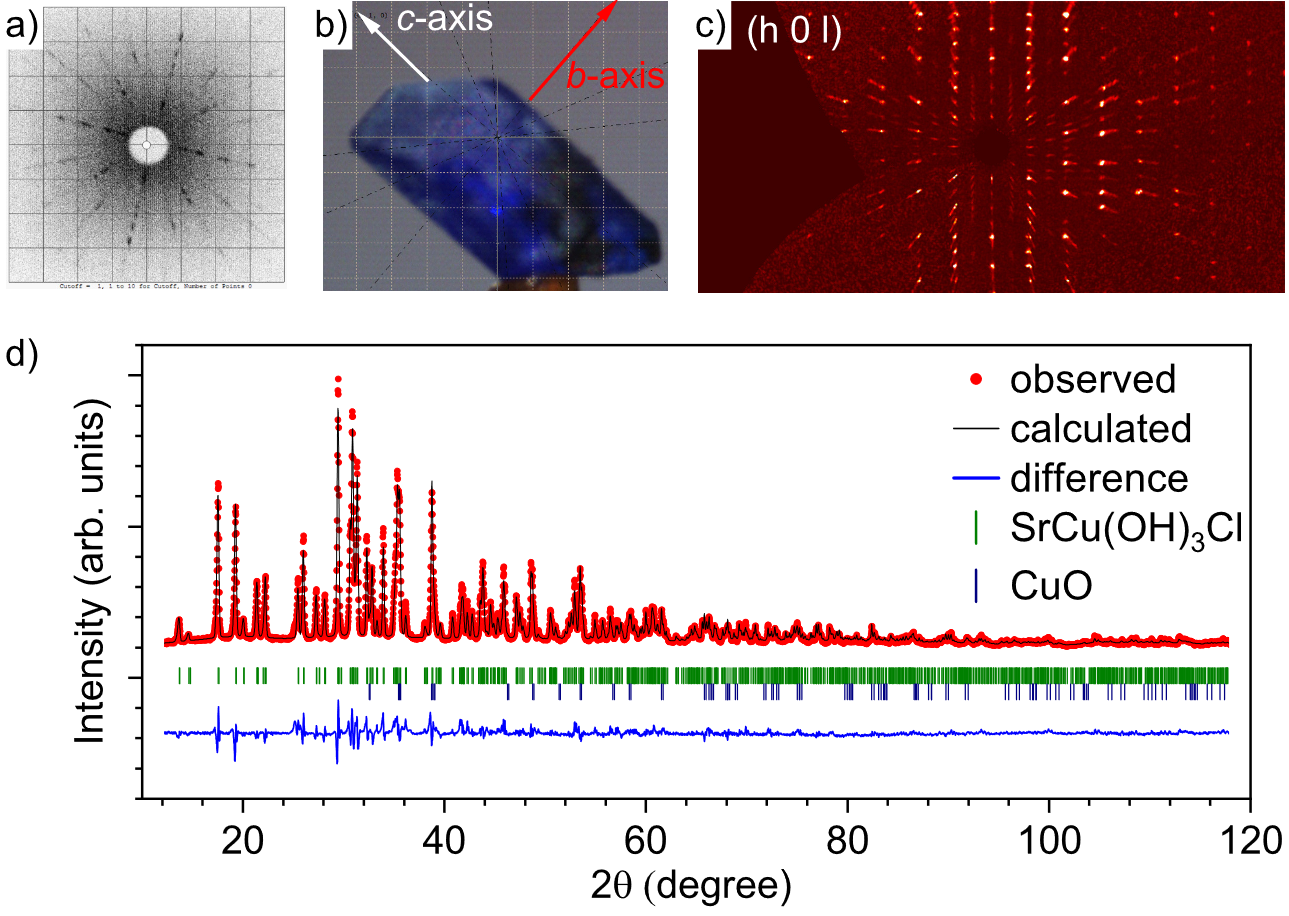}
\caption{x-ray diffraction of SrCu(OH)$_3$Cl single crystals: a) Laue diffraction along the $a$-axis b) Picture of a single crystal with axis indication c) Single crystal diffraction map of the (h0l) plane and d) powder diffraction pattern of \RK{crushed crystals}.} 
\label{xrd}
\end{figure}

Crystals were checked with single crystal x-ray diffraction \RK{(XRD)} and oriented with Laue diffraction for the susceptibility measurements (see Fig. \ref{xrd}(a)). As reported in \cite{Zhu2014} the system crystallizes in $Pmn$2$_1$ (space group no. 31). The lattice parameters extracted from the single crystal x-ray diffraction data amount to $a$ = 15.984(2) \AA, $b$ = 6.3905(11)~\AA, and $c$ = 6.4694(17)~\AA, in agreement with those given by Zhu \textit{et al.}. The  refined atom positions  are compiled in Table {\color{blue} I}. The powder diffraction patterns revealed a minute CuO impurity. \RK{Some crystals were crushed for powder x-ray powder diffraction.}

\begin{table}[h]
\centering
\caption{Refined \RK{atom} positional coordinates of SrCu(OH)$_3$Cl in the orthorhombic space group $Pmn2_1$ (no. 31), extracted from single crystal XRD.} \label{posit}
\begin{tabular}{llllll}
\\\hline
Atom & x & y & z & Uiso ($\textrm{Å}^{2}$) & Occ.\\\hline
Sr1 & 0.5 & 0.3352(4) & 1.0722(5) & 0.0065(6) & 1\tabularnewline
Sr2 & 0.17326(9) & 0.9289(3) & 0.2429(4) & 0.0091(5) & 1\tabularnewline
Cu1 & 0.39912(13) & 0.7980(3) & 0.2488(5) & 0.0074(6) & 1\tabularnewline
Cu2 & 0.5 & 0.4702(5) & 0.5678(7) & 0.0071(7) & 1\tabularnewline
Cl1 & 0.2636(3) & 0.6757(7) & -0.0758(10) & 0.0146(11) & 1\tabularnewline
Cl2 & 0 & 1.0588(12) & 0.1321(14) & 0.0177(17) & 1\tabularnewline
O1 & 0.4166(8) & 0.516(2) & 0.353(2) & 0.007(3) & 1\tabularnewline
O2 & 0.3842(9) & 1.065(2) & 0.116(2) & 0.009(3) & 1\tabularnewline
O3 & 0.3043(9) & 0.838(2) & 0.431(3) & 0.009(3) & 1\tabularnewline
O4 & 0.4129(8) & 0.394(2) & 0.758(3) & 0.009(3) & 1\tabularnewline
O5 & 0.5 & 0.747(3) & 0.092(3) & 0.007(4) & 1\tabularnewline
\hline
\end{tabular}
\end{table}

\begin{figure}[h]
\centering
\includegraphics[width=1.0\columnwidth]{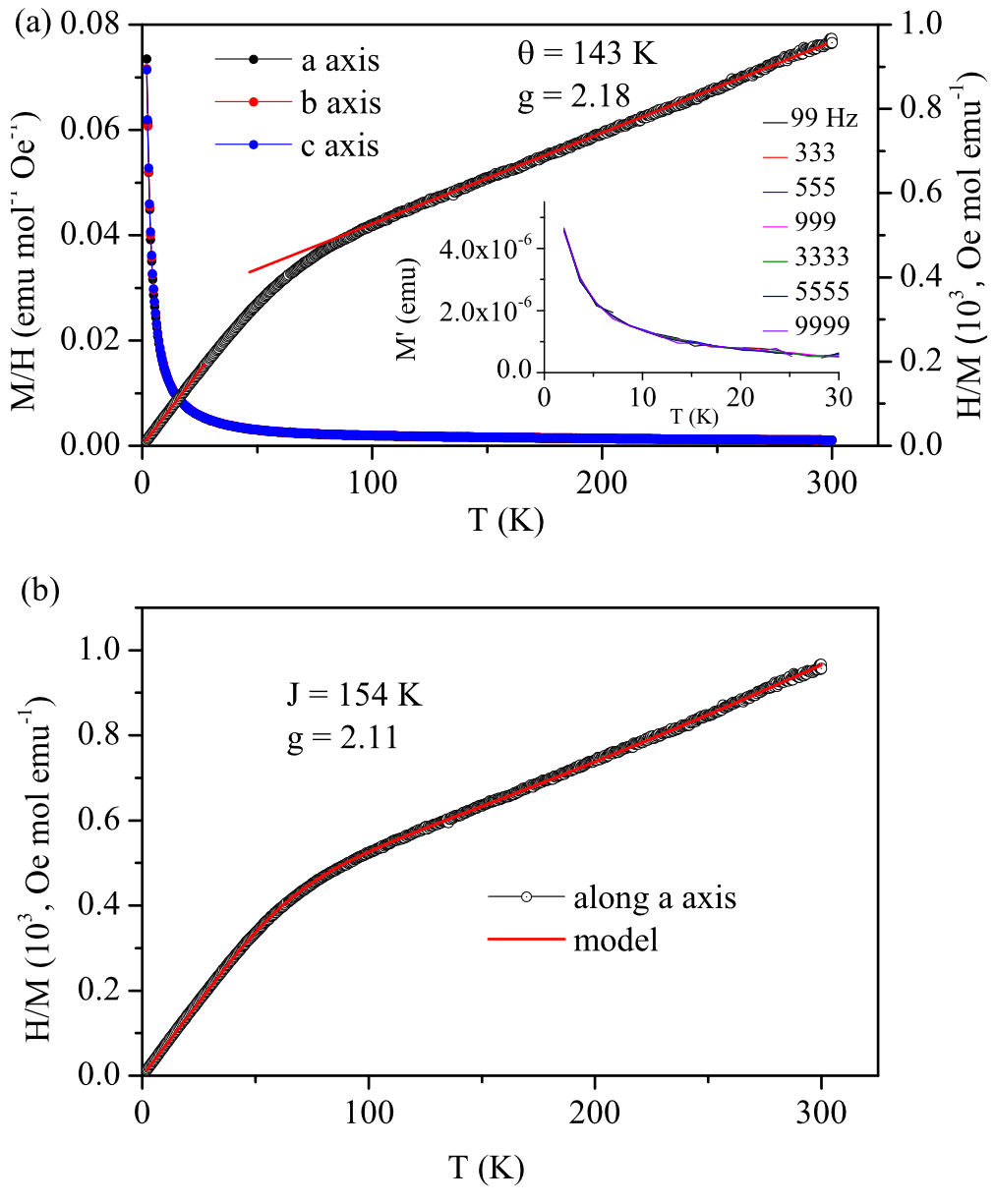}
\caption{(a) Temperature dependence of dc magnetic susceptibility along three crystallographic direction measured at $\mu_{\rm 0}$H = 1 T (notably the three curves lie nearly indistinguishable atop each other). The right panel shows the temperature variation of the inverse susceptibility along \RK{the $a$-axis}, where the red lines are the Curie-Weiss fits at high and low temperatures. Inset shows the frequency dependence of ac susceptibility at low temperatures. (b) shows the fitting of the inverse susceptibility assuming isotropic exchange interaction between spins on an equilateral triangle.}
\label{susceptibility}
\end{figure}

\subsection{\label{sec:level}dc magnetic measurements}
In Fig. {\color{blue}\ref{susceptibility}(a)}, we show the temperature dependence of the magnetic susceptibility ($M/H$) of SrCu(OH)$_3$Cl along the three crystal axes measured in \RK{an} external field of $\mu_{\rm 0}$H = 1 T. The measurements were carried out in the zero field cooled measurement protocol. The susceptibility monotonically increases with decreasing temperature. It does not show any anomaly indicative of a phase transition down to the lowest measured temperature of $T$ = 2 K, which is in accordance with the earlier report \RK{by Zhu \textit{et al.}} {\color{blue}\cite{Zhu2014}}.  In the right panel of  Fig. {\color{blue}\ref{susceptibility}(a)}, we plot the temperature variation of the inverse susceptibility, $\chi^{-1}(T)$ along the crystallographic $a$-direction in the temperature range between $T$ = 2-300 K. At higher temperatures, $\chi^{-1}(T)$ exhibits a linear $T$-dependence, which changes its slope between $T$ = 50-80 K and again becomes linear at low temperatures. The susceptibility above $T$ $\sim$ 80 K can be fitted well using Curie-Weiss law with Weiss temperature of $\theta_{CW}$ = -143 K and effective moment of $\mu_{eff}$ = 1.85 $\mu_B$ (g = 2.18). On the other hand at lower temperatures, the linear fit of $\chi^{-1}(T)$ yields the Curie-Weiss temperature is $\theta_{CW}$ = 0.04 K. The Curie-Weiss law at high temperatures proves a rather large intra-trimer antiferromagnetic interaction ($J\sim 10^2$ K) as already concluded by Zhu \textit{et al.} The inter-trimer antiferromagnetic interaction is extremely small as suggested by the small Curie-Weiss temperature obtained from the data at lower temperatures.

\begin{figure}[h]
\centering
\includegraphics[width=1.0\columnwidth]{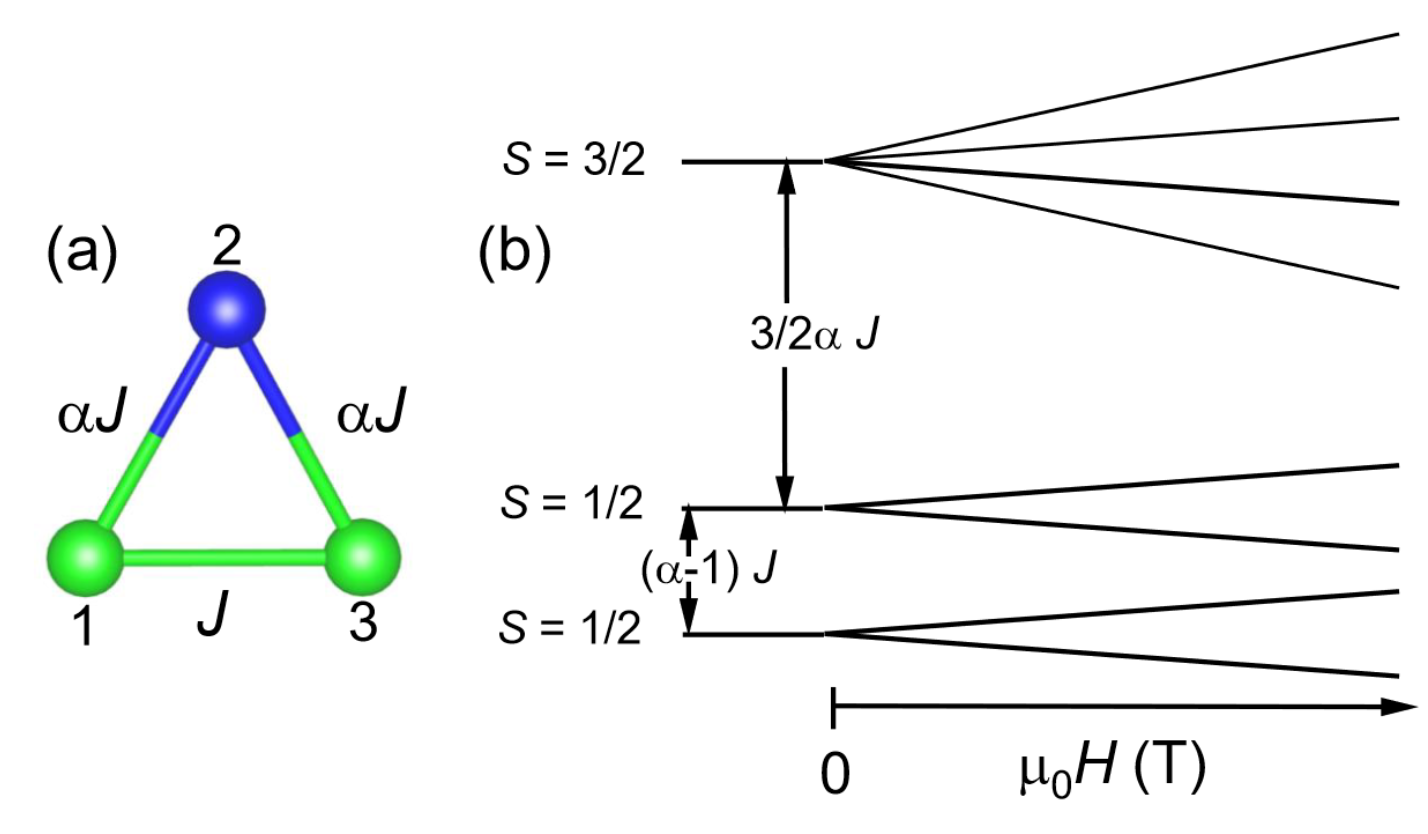}
\caption{(a) Definition of the spin exchange parameters. $\alpha J$ and $J$, for the Cu isosceles triangles in SrCu(OH)3Cl. Blue and green spheres denote the different Cu atom sites. (b) Energy level diagram of the isosceles triangle (not to scale). In case of $\alpha$ = 1 the two doubly degenerate \RK{$S$ = 1/2} levels are degenerate.} \label{Isosceles}
\end{figure}

Zhu \textit{et al.} have fitted the full temperature dependence of the magnetic susceptibility to a model assuming identical exchange parameters between the three spin $S$ = 1/2 entities. However, the crystal structure data of SrCu(OH)$_3$Cl revealed two different Cu atoms, rather indicating that a description in terms of  an isosceles triangle with  two identical exchange parameters ($\alpha\,J$) and  one exchange ($J$) may be appropriate to model the the full temperature dependence of the magnetic susceptibility (see Fig. \ref{Isosceles}).

Haraldsen \textit{et al.} have calculated the temperature dependence of the magnetic susceptibility of an isosceles triangle describe by the Hamiltonian given by Eq. (1). {\color{blue}\cite{Haraldsen2005}} It is given by


\begin{equation}
\frac{4k_{\rm B}T\chi}{g^2\mu_B^2} =  \frac{10+exp(3\alpha\,J/2k_{\rm B}T) +exp([2+\alpha]\,J/2k_{\rm B}T)}%
{2+exp(3\alpha\,J/2k_BT) + exp([2+\alpha]\,J/2k_{\rm B}T)},
\label{triangle}
\end{equation}
which for $\alpha$ = 1 converts to the equation used by Zhu \textit{et al.} to fit their data.
Fig. {\color{blue}\ref{susceptibility}(b)} displays our fit of the magnetic susceptibility measured with magnetic field applied along the $a$-axis indicating an exchange parameter
\begin{equation*}
J = 154 ~{\rm K}.
\end{equation*}
We note that our value is substantially smaller than that given by Zhu \textit{et al.} (233 cm$^{-1} \approx$ 335 K). Interestingly the fits converge to $\alpha \approx$ 1, suggesting a symmetric \RK{i.e. equilateral} trimer, despite the presence of two different Cu atoms and different bonding distances and Cu -- O -- Cu bonding angles.
Model calculations with $J$ = 154~K and $\alpha$ = 1 yield  a Curie-Weiss temperature of -133~K, very close to the result of the Curie-Weiss fit of our high temperature data.

Assuming an additional temperature independent term, $\chi_0$ due to core diamagnetism and van Vleck contribution, the $\chi^{-1}-T$ curve can be approximated very well by (Eq. \ref{triangle}).  From the fitting, we obtain $J$ = 154 K, and $g$ = 2.11 and $\chi_0$ = -60 $\times$ 10$^{-6}$~emu/mole Oe. The diamagnetic contribution estimated for SrCu(OH)$_3$Cl is $\chi_{dia}$ $\sim$ -90$\times$10$^{-6}$ emu/mole Oe {\color{blue}\cite{Bain2008}}, which gives an approximate estimation of the van Vleck contribution of about $\chi$ (van-Vleck) = 30$\times$10$^{-6}$ emu/mole Oe. In the inset of Fig. {\color{blue}\ref{susceptibility}(a)}, we have also shown the temperature dependence of the real part of the ac susceptibility in the low temperature region. The ac susceptibility is independent of frequency in the probing frequency and temperature range.

\begin{figure}[h]
\centering
\includegraphics[width=1.0\columnwidth]{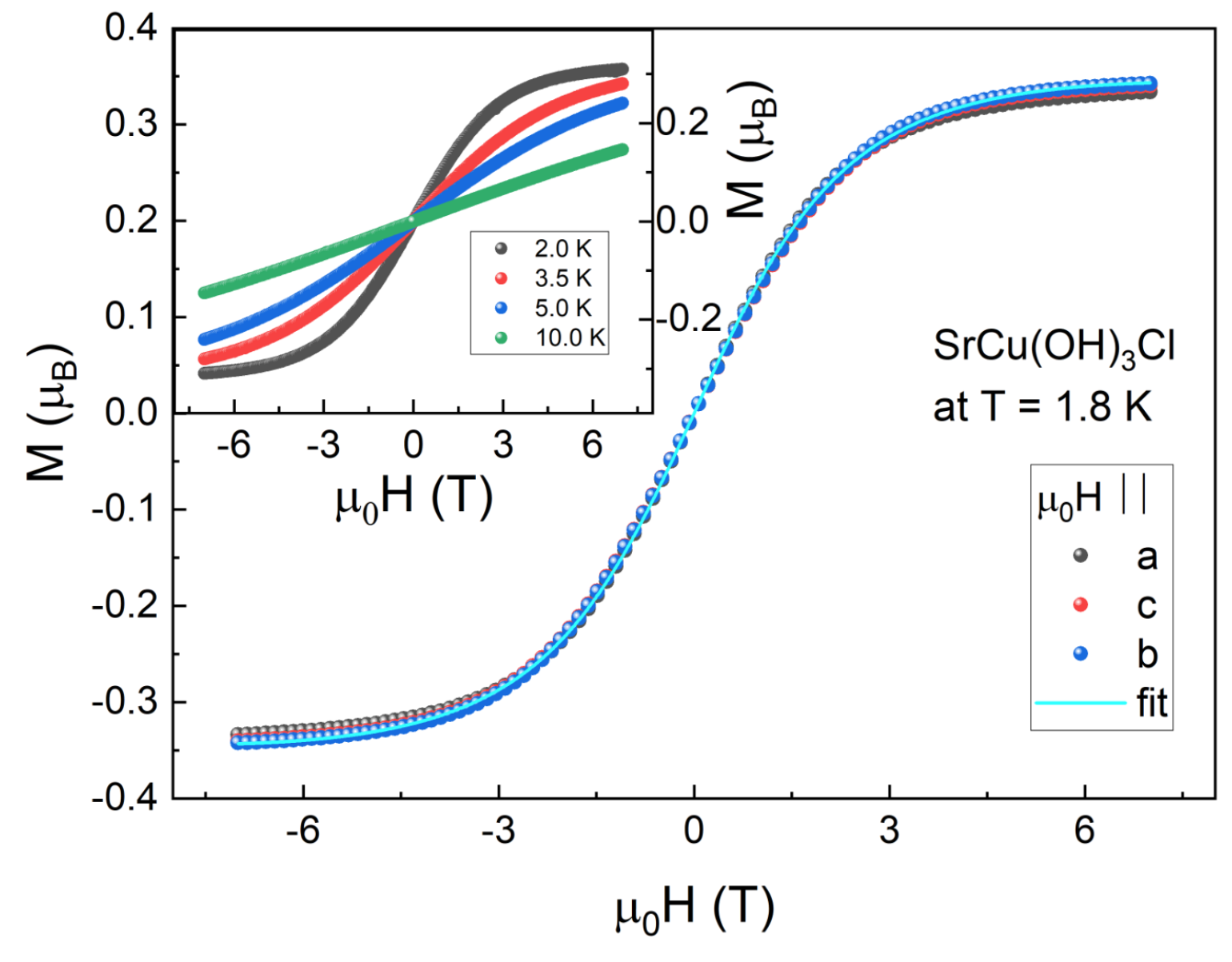}
\caption{The main \RK{panel} shows the magnetization versus field applied along all three crystallographic directions at 1.8 K. The solid bright blue line represents a fit with a Brillouin function for $S$ = 1/2 and $g$ = 2.11. The inset displays the isothermal magnetization curves for various temperatures.} \label{Brill}
\end{figure}

Fig.  {\color{blue}\ref{Brill}} displays the isothermal magnetization collected at 1.8 K with the magnetic field aligned along the crystal needle ($c$-axis). Measurements with $a$-axis, $b$-axis and $c$-axis orientation are identical within error bars. 

In the main panel of the Fig. {\color{blue}\ref{Brill}}, we show the magnetic field dependence of the dc magnetization (isothermal $M$-$H$) curves along the three crystal axes at temperature $T$ = 1.8 K.  Note that, along all three directions the $M$-$H$ curves reaches to a saturation value of the magnetic moment of  $\sim$0.33 $\mu_B$/Cu atoms, i.e. 1/3 of $M_s$ \RK{per formula unit}. 
The magnetization can be easily fitted with a Brillouin function for S=1/2 and $T$ =1.8 K.

At high magnetic fields the magnetization per Cu trimer saturates to a magnetic moment of 1/3 $\mu_{\rm B}$ per formula unit. The field dependence follows very well the Brillouin function for one isolated spin $S$ = 1/2 \RK{entity}, indicating that each Cu triangle is antiferromagnetically coupled and has a ground state of an effective spin 1/2 and is magnetically well isolated from its neighboring triangles.

\subsection{\label{sec:level}Specific heat measurements}
\begin{figure}[h]
\centering
\includegraphics[width=1.0\columnwidth]{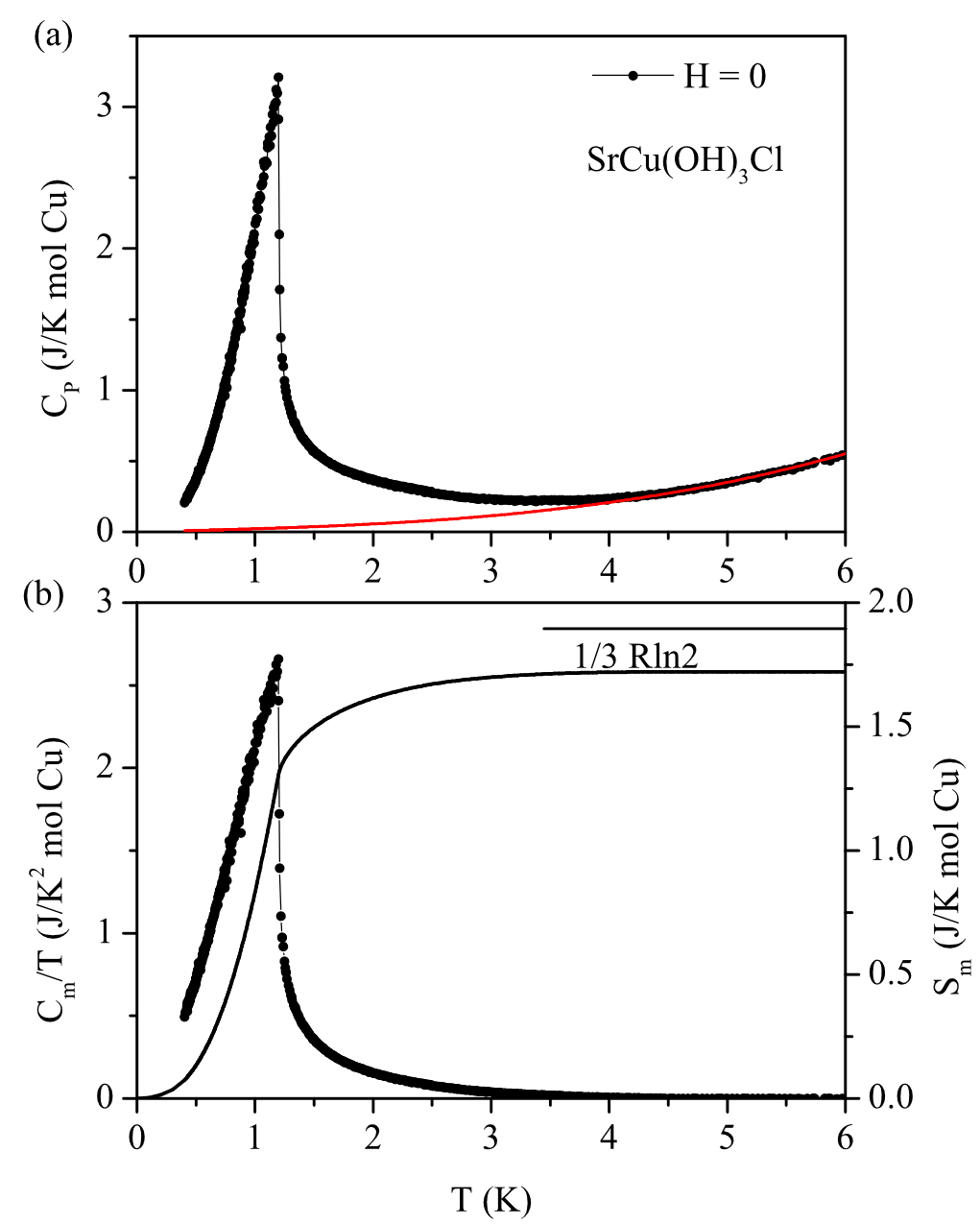}
\caption{(a) Temperature dependence of the total specific heat at $H$ = 0. It \RK{reveals} a $\lambda$-anomaly at \RK{$T_N$ = 1.2~K}. The red line gives the approximation from its \RK{phonon} contribution \RK{to the heat capacity}. (b) the left panel shows the temperature dependence of $C_m/T$, across $T_N$, where $C_m$ is the magnetic heat capacity. The right panel displays the magnetic entropy released across the transition. }
\label{HC0}
\end{figure}

To further investigate the magnetic ground state, we have carried out temperature and magnetic field dependent  specific heat measurements of SrCu(OH)$_3$Cl down to 400 mK. The total specific heat as a function of temperature at $H$ = 0 is plotted in Fig. {\color{blue}\ref{HC0}(a)}.  The zero-field specific heat shows a $\lambda$-type anomaly at $T$ = 1.2 K, and a weak broad hump-like feature at higher temperatures centered around $T$ = 1.9 K. The $\lambda$-type feature proves that the system undergoes a second order phase transition below $T_{rm N}$ = 1.2 K, apparently of magnetic origin due to some very weak exchange interaction between the 'compound' spins of the Cu triangles.
The broad shoulder above the $\lambda$ anomaly can be ascribed to short range spin fluctuations preceding the long range magnetic order at $T_{\rm N}$. 

\begin{figure}[h]
\centering
\includegraphics[width=1.0\columnwidth]{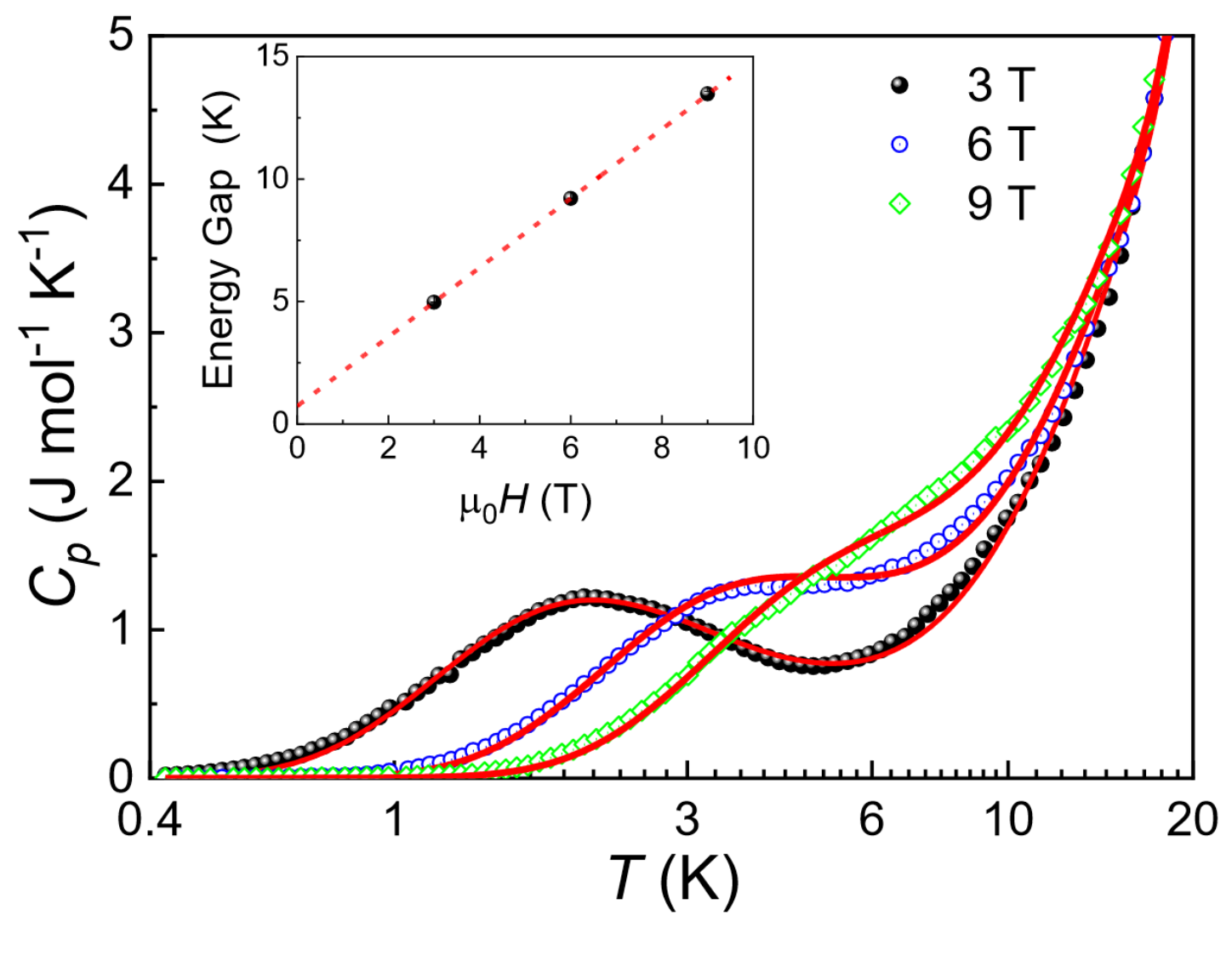}
\caption{Specific heat curves of SrCu(OH)$_3$Cl measured in magnetic fields starting from 3 T. The (red) solid lines are fits with the heat capacity expected for a two-level system including a phonon term as described in detail in the text. The inset displays the level-splitting as a function of the magnetic field.}
\label{Schottky}
\end{figure}

To obtain the magnetic heat capacity contribution we subtracted a phonon part from the total heat capacities. The phonon contribution shown by the solid red line in Fig. {\color{blue}\ref{HC0}(a)} was determined by fitting a $T^3$ power law to the heat capacities above $\sim$5 K and extrapolating it to $T \rightarrow$ 0 K. 
The magnetic entropy displayed in Fig. {\color{blue}\ref{HC0}(b)} was calculated by integrating the magnetic heat capacity, $\int({C_{\rm m}/T})dT$.
It saturates to a value very close to $\frac{1}{3}$$R$ln(2), where $R$ is the molar gas constant.
The $S_{total}$ = 1/2 ground state of an equilateral triangle cluster is fourfold degenerate. A complete lifting of the degeneracy e.g. by a small asymmetry of the triangle exchange (i.e. $\alpha \neq 1$) and an inter trimer exchange interaction the entropy would add up to \RK{2}$R$\,ln(2). For $\alpha = 1$, however, the entropy remains at $R$ln(2), since  the $S_{total}$ = 1/2 degeneracy is not lifted  and the ground state splits into two states, each with  twofold degeneracy {\color{blue}\cite{Haraldsen2005}}.
In addition to the magnetic susceptibility and the isothermal magnetization measurements that found saturation at a value of $\sim$1 $\mu_{\rm B}$ per Cu triangle, the magnetic entropy therefore also supports $\alpha$ = 1, i.e. a 'magnetically symmetric' equilateral triangle despite the structural asymmetry.
\begin{figure}[h]
\centering
\includegraphics[width=1.0\columnwidth]{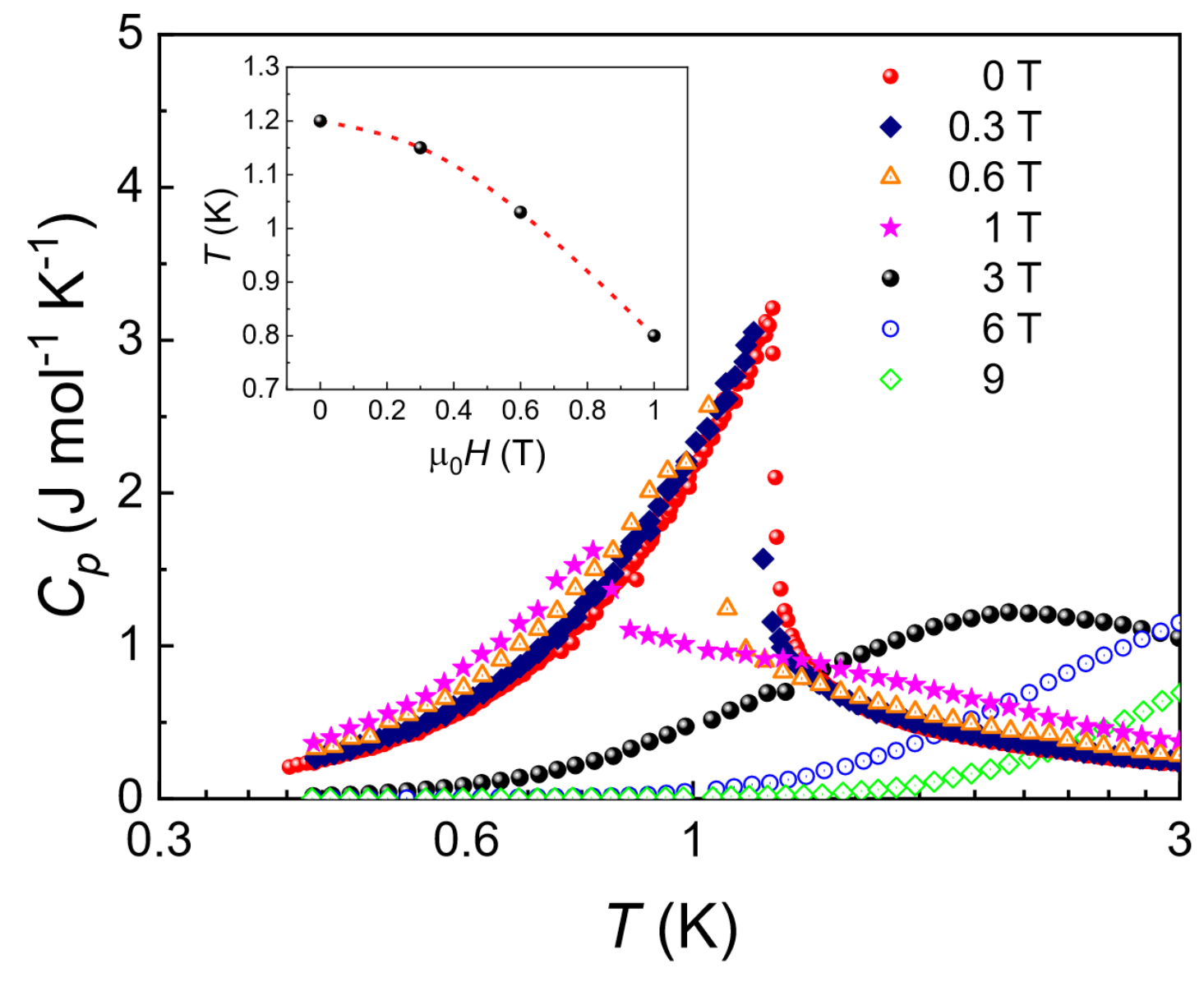}
\caption{a) Magnetic field dependence of \RK{the} specific heat of SrCu(OH)$_3$Cl applied along the $a$-direction. The $\lambda$ anomaly associated with antiferromagnetic transition is gradually suppressed with increasing magnetic field (see upper inset). At higher field an additional broad anomaly appears which shifts to higher temperatures with field.}
\label{C-field-dependent}
\end{figure}
The broad heat capacity anomaly \RK{appearing at higher magnetic fields} can be successfully fitted by assuming a two-level Schottky system with an energy gap $\Delta$ and a degeneracy of two for each level.
The red solid lines in  Fig. \ref{Schottky} displays the heat capacities fitted to a  system using Eq. (\ref{SchoFits}) including a phonon contribution. 
\begin{equation}
C_p=R\cdot \frac{1}{3}\cdot \frac{g_1}{g_2}\cdot (\Delta/k_{\rm B}T)^2\cdot e^{\Delta/k_{\rm B}T}/ (1+e^{\Delta/k_{\rm B}T})^{2}+C_{\rm {phon}},      
 \label{SchoFits}
 \end{equation}
\RK{$g_1$ and $g_2$ are the degeneracies of the two levels}
for the phonon heat capacity contribution, $C_{\rm phon}$, we used a polynomial description according to
\begin{equation}
C_{\rm phon} = \beta T^3 +\gamma T^5 + \delta T^7     
\label{Cphon}
\end{equation} 

The energy difference $\Delta$ of the two levels increases linearly with magnetic field as shown in detail in the inset in  Fig. {\color{blue}7} with an intersection of 0.73 K for vanishing magnetic field, indicating an internal exchange field of $\sim$ 1 T causing the long-range antiferromagnetic ordering at 1.2 K.

\subsection{\label{sec:level}Nuclear Magnetic Resonance}
\begin{figure}[h]
	\centering
	\includegraphics[width=1\columnwidth]{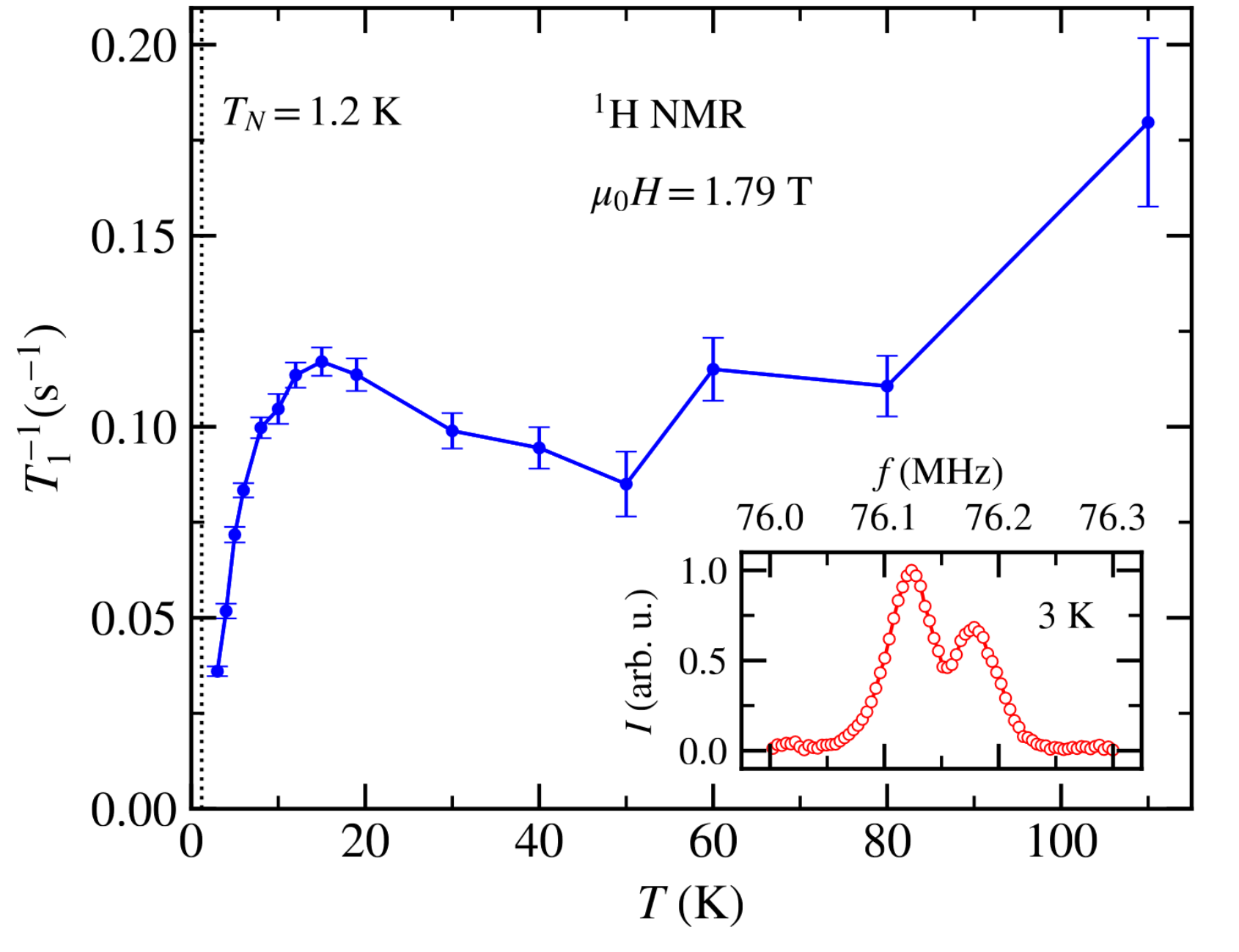}
	\caption{Temperature-dependent spin-lattice relaxation rate probed by $^1$H NMR. Inset: the spectrum at 3~K exhibits similar line splitting into two peaks as related single-crystalline Cu$^{2+}$ hydroxide systems {\color{blue}\cite{Wang2023}}.} 
	\label{Fig:1H_NMR_T1_SrCuOH3Cl}
\end{figure}

Temperature dependent $^1$H NMR was used to probe the paramagnetic state of SrCu(OH)$_3$Cl. The NMR spectrum collected at 3~K in an external magnetic field of 1.79~T, shown in the inset of Fig.~\ref{Fig:1H_NMR_T1_SrCuOH3Cl} consists of two resonance lines at 76.12~MHz and 76.18~MHz and a FWHM of $\sim$40~kHz. The intensity ratio of the two resonance line amounts to $\sim$3 : 2, reflecting the five in-equivalent proton positions in the crystal structure of SrCu(OH)$_3$Cl.\cite{Zhu2014}

The temperature-dependent spin-lattice relaxation rate, $T_1^{-1}$, of the protons in the hydroxyl groups  was analyzed to probe the paramagnetic state and the magnetic fluctuations above the magnetic ordering at $\sim$1.2~K. Fig.~\ref{Fig:1H_NMR_T1_SrCuOH3Cl} displays $T_1^{-1}$.   Upon cooling below 100~K $T_1^{-1}$ decreases and exhibits a local minimum at around $\sim$50~K. The minimum  coincides with the change in slope of inverse susceptibility (see) Fig.~\ref{susceptibility}) corroborating the  picture of the thermal population of isolated spin trimers spin on the Cu$^{2+}$ triangles.
At lower temperatures, $T_1^{-1}$ develops a broad maximum at around 15~K, below which the relaxation rate rapidly drops. We assign this decrease to the reduction of magnetic fluctuations on approaching the long-range antiferromagnetic ordering at 1.2~K. 

\subsection{\label{sec:level}Electron Spin Resonance}
\begin{figure}[t]
\centering
\includegraphics[width=1.0\columnwidth]{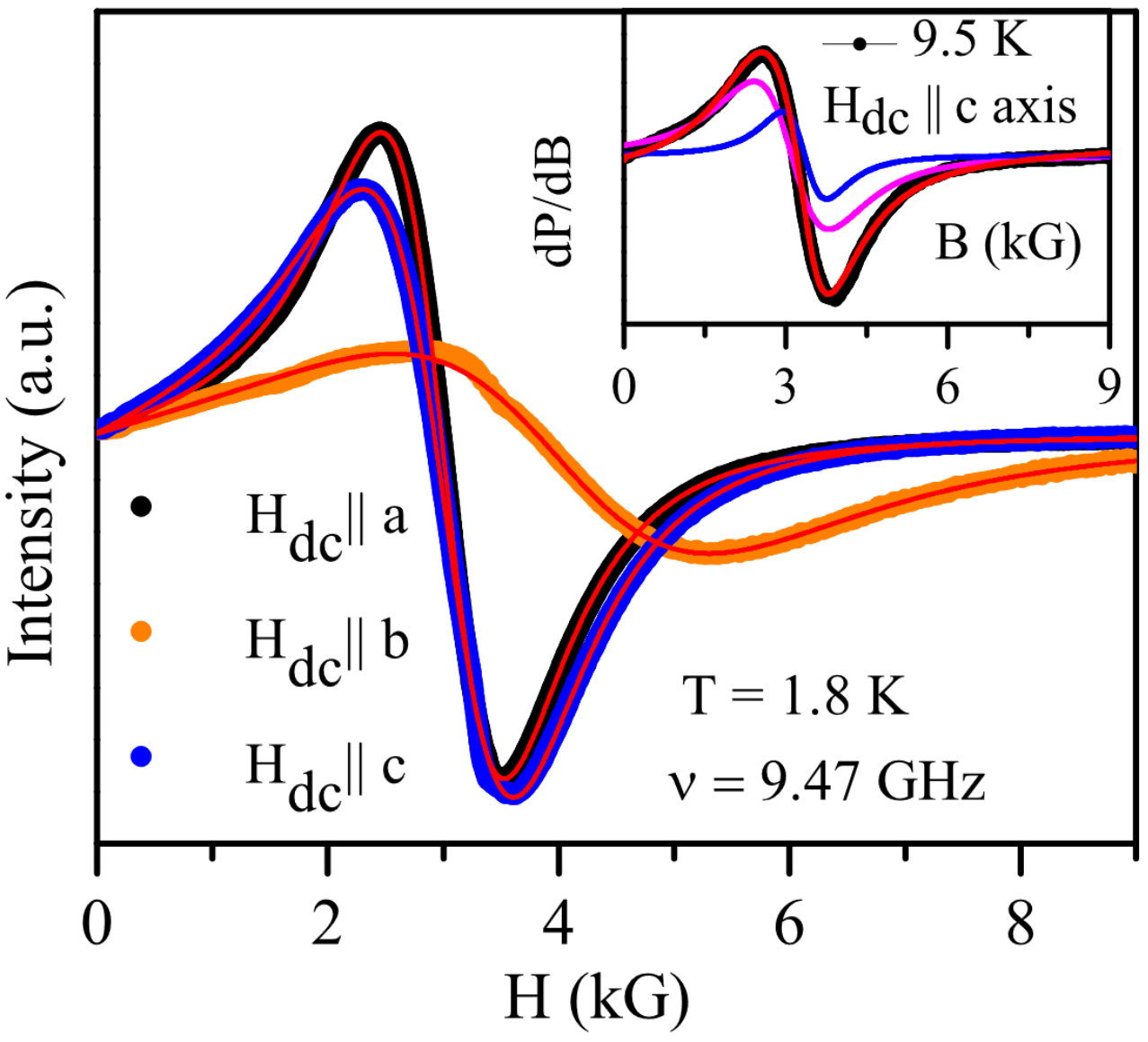}
\caption{The main panel shows the ESR resonance of SrCu(OH)$_3$Cl along the three crystal axes at $T$ = 1.8 K. The red lines are fits of the ESR lines using Eq. (\ref{ESReq}). Inset: the fitting of the ESR line recorded at $T$ = 9.5 K along $H_{dc}$$ || $c axis using two Lorentzian lines as described in detail in the text.}.
\label{EsrSpectra}
\end{figure}

In Fig. \ref{EsrSpectra}, we show  representative ESR spectra along the three crystal axes recorded at $T$ = 1.9 K. As \RK{the} temperature increases, the ESR signal becomes broader and the intensity reduces dramatically, which make \RK{it} difficult to reliably record any ESR signal above $T$ = 35 K. In all directions, the ESR spectra are well described by the field derivative of a single Dysonian line. Since the ESR line width is large and of the same order of magnitude as the resonance field, we have to incorporate the contributions from both the right circular and left circularly polarized components contained in the linearly polarized microwave excitation \cite{Caslin2014,Ivanshin2000}. Hence, the resonance at the reversed magnetic field -$H_{0}$ has to be included into the fitting equation, which is then given by

\begin{equation}
\frac {dP}{dH} \propto \frac {d}{dH} \bigg[{{\frac{W+\sigma(H-H_0)}{H_0^2 + (H-H_0)^2}}+{\frac{W+\sigma(H+H_0)}{H_0^2 + (H+H_0)^2}}}\bigg].
\label{ESReq}
\end{equation}

Here, $H_0$ and $W$ are the resonance field and the linewidth, respectively. The parameter $\sigma$ in Eq. (\ref{ESReq}) denotes the dispersion-to-absorption ratio, which describes the asymmetry of the ESR line-shape.  Such \RK{asymmetric ESR lines are} generally observed in metallic samples where the skin effect due to large conductivity of the material causes the electric and magnetic fields of the microwave out of phase.\cite{Gambke1983} Therefore, an asymmetric ESR line-shape is \textit{per se} not expected in an insulating compound like SrCu(OH)$_3$Cl.  In case of low dimensional spin systems, a finite $\sigma$ may  arise in magnetic insulators due to admixture of non-diagonal elements of the dynamic susceptibility to the signal, which distorts the ESR signal with a large value of the linewidth {\color{blue}\cite{Benner1983}}. Such behaviour has been observed in one-dimensional spin chain compounds with strong spin-orbit coupling {\color{blue}\cite{Benner1983,Seidov2001}}. Moreover, in these systems the ESR spectra is symmetric at high temperatures and the distortion parameter, $\sigma$ gradually increases as the system approaches $T_N$ from the high temperatures side. As we will discuss later, the temperature variation of $\sigma$ in  SrCu(OH)$_3$Cl is more complex. One possibility of such asymmetric ESR absorption might be a convolution of multiple ESR absorption \RK{lines}. In SrCu(OH)$_3$Cl, there are two in-equivalent Cu atoms within a spin-trimer. In the inset of Fig. \ref{EsrSpectra}, we show the modeling of the ESR line at $T$ = 9.5~K measured along $H_{dc}$ $||$ $c$-axis assuming two Lorentzian lines with 2:1 intensity ratio and g = 2.15(3) and 2.02(3) and linewidths of 1275(30) and 692(20) G. However, due to large linewidth, and absence of clear splitting between the two resonance lines, the significance of such fits is limited. In this context, it may also be mentioned that the ESR spectra of finite size systems such as XXZ spin chain with finite lengths or molecular magnets shows a double-peak structure at high temperatures which strongly differs from usual Lorentzian \RK{line-shape} {\color{blue}\cite{ Shawish2010}}. The separation of the peaks of the double-peak structure also vanishes inversely with the system size {\color{blue}\cite{Ikeuchi2017}}. Therefore, the origin of the asymmetry in the present system is not clear.\\

\begin{figure}[htp!]
\centering
\includegraphics[width=1.0\columnwidth]{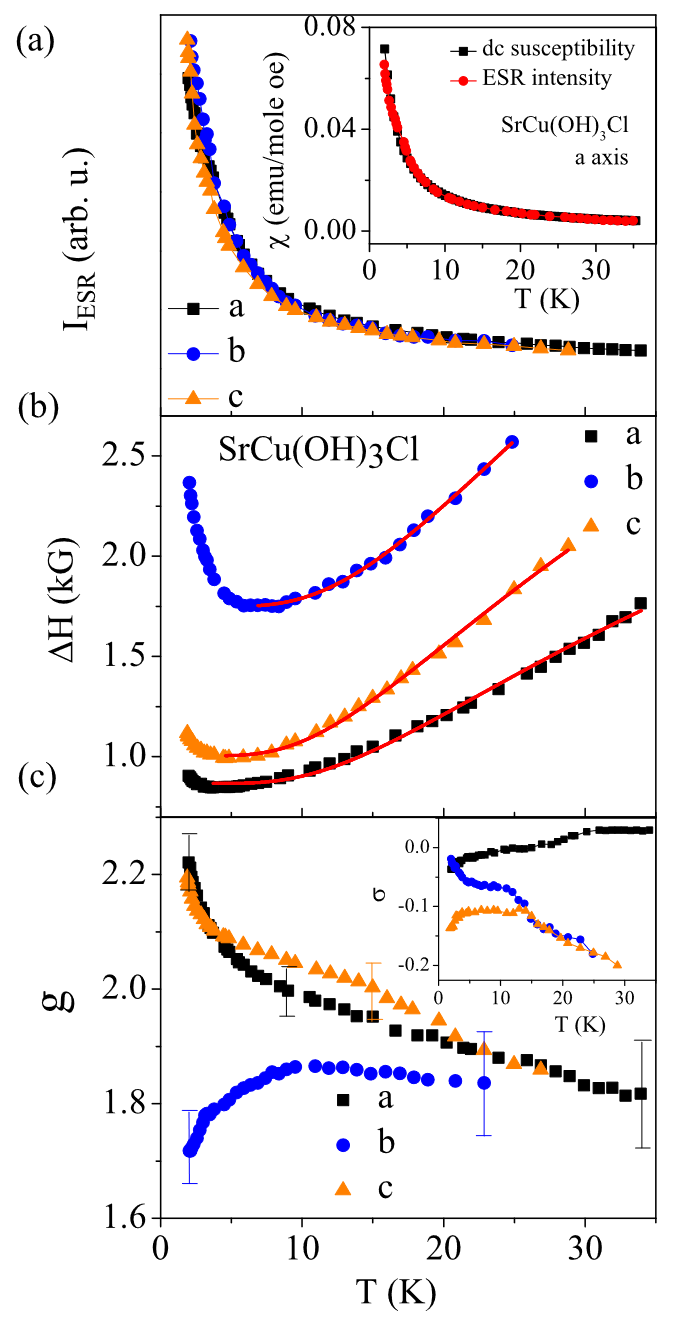}
\vspace{-0.5cm}
\caption{Temperature dependence of ESR parameters of SrCu(OH)$_3$Cl. (a) The main panel shows the total intensity of the ESR signal along three different crystallographic directions. Inset: Comparison of the dc susceptibility measured in a SQUID magnetometer and the ESR intensity. (b) the temperature variation of ESR line width along three directions. The red lines correspond to the fits of the high temperature region using Eq. (\ref{delta}). (c) The main panel shows the temperature variation of the apparent g-factor obtained from the ESR absorption. Inset: variation of the dispersion parameter, $\sigma$ in Eq. (\ref{ESReq}) as a function of temperature.}
\label{EsrIntens}
\end{figure}

Nonetheless, to further understand the ESR properties of SrCu(OH)$_3$Cl, we have fitted the spectra measured along all three directions using Eq. (5), (see in Fig.~\ref{EsrSpectra}). The ESR intensity, $I_{\rm ESR}$ which is obtained by the double integration of the ESR spectra is shown in Fig.~\ref{EsrIntens}{\color{blue}(a)}.   Consistent with the dc magnetization measurements, the intensity $I_{ESR}$ along all three direction monotonically increases with decreasing  temperature. The intensity does not exhibit any signature of spin freezing or phase transition down to $T$ = 1.8 K. In the inset of Fig.~\ref{EsrIntens}(a), we have compared the dc susceptibility ($\chi_{\rm dc}$) with the ESR intensity ($I_{\rm ESR}$) below $T$ = 35~K along the crystallographic $a$-axis. To compare the intensity $I_{ESR}$ with $\chi_{dc}$, $I_{ESR}$ has been normalised to the dc susceptibility data at $T$ = 35~K.    

\begin{figure}[h]
\centering
\includegraphics[width=1.0\columnwidth]{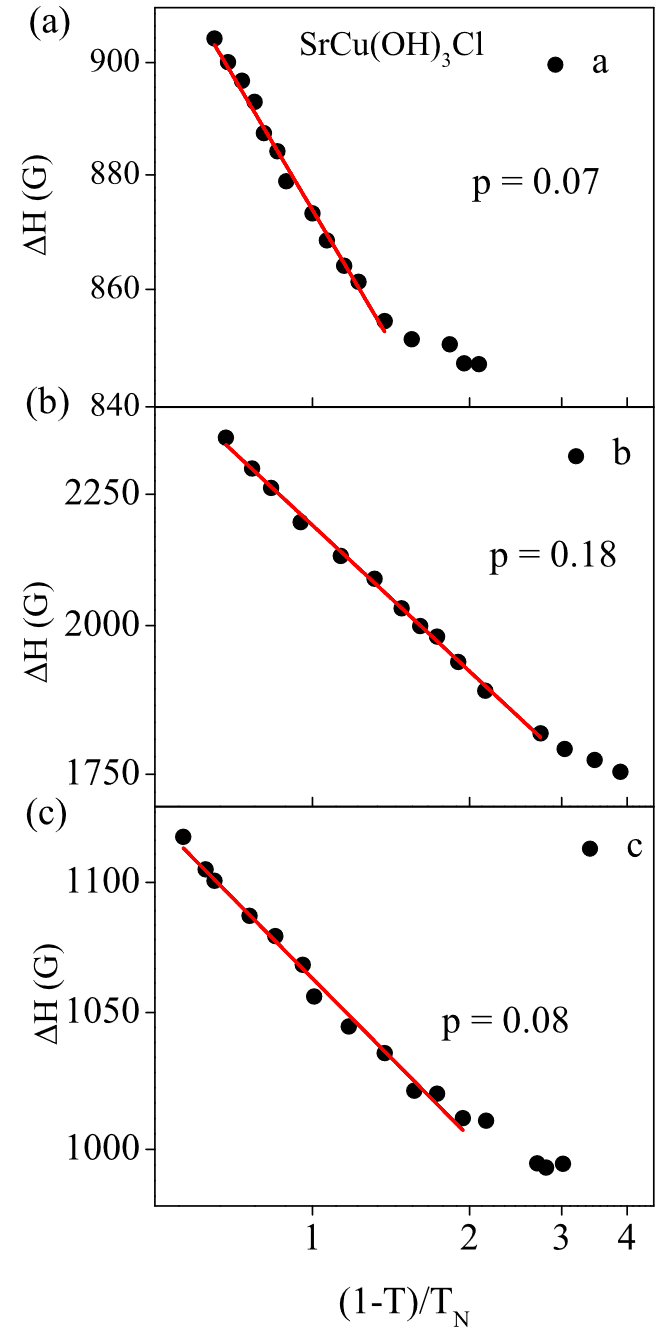}
\caption{Double logarithmic plot of $\Delta H$ and (1-T)/$T_N$ along three crystallographic directions, with $T_N$ =~1.2 K, which is obtained from the specific heat measurements at $H$ = 0. The red lines show the linear fit to the data points. The slopes of the linear fit, which gives the exponent are also mentioned.}
\label{EsrCritical}
\end{figure}

\begin{figure}[h]
\centering
\includegraphics[width=1.0\columnwidth]{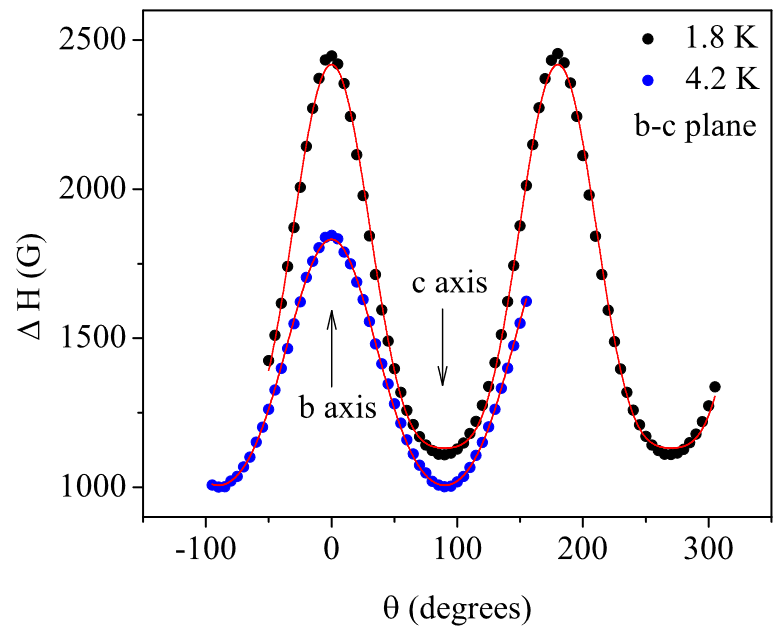}
\caption{Angle dependence of the linewidth in the $b$-$c$ plane at $T$ = 1.8 K, and 4.2 K. The red line is the fitting of the data as described in detail in the text.}
\label{EsrAngle}
\end{figure}

\subsubsection{\label{sec:level}Critical-point anomalies in ESR linewidth} 
In Fig.~\ref{EsrIntens}{\color{blue}(b)}, we show the temperature variation of the ESR linewidth determined with the magnetic field aligned parallel to the crystallographic $a$, $b$, and $c$ axes. As the temperature is reduced, the linewidth initially decreases, passes through a shallow minimum in the temperature range between $T$ = 4-10~K, and then increases rapidly on approaching  $T_N$. Particularly, the linewidth along $H_{dc}$ $||$ $b$ axis is much larger compared to $H_{dc}$ $||$ $a$ and the $c$ axes. Moreover, at low temperatures the linewidth exhibits a stronger divergence along $b$-axis, as will be discussed later. At high temperatures the temperature dependence of the linewidth along all three crystal directions can be fitted well using the following the empirical equation {\color{blue}\cite{Heinrich2003}}:
\begin{equation}
\Delta H = \Delta H_{\infty}+ \Delta H_{act} \cdot exp(-E_g/T).
\label{delta}
\end{equation}
Here, $\Delta H_{\infty}$ denotes the high-temperature asymptotic contribution due to pure spin-spin relaxation, and the last term describes a thermally activated contribution $\Delta H_{act}$ characterized by an energy gap $E_g$. The values of $\Delta H_{act}$ and the energy gap $E_g$ along three directions considering both of them as free parameters are listed in the Table II. 

\begin{table}[h]
\centering
\caption{Parameters for ESR linewidth}
\begin{tabular}{llll}
\\\hline
Directions  & \textit{a} & \textit{b} & \textit{c} \\\hline 
$\Delta H_{act}$ (kG)  & 2.7 & 6.0 & 4.2 \tabularnewline
$E_g$ (K)  & 40.3 & 48.5 & 41 \tabularnewline
\hline
\end{tabular}
\end{table}


The ESR linewidth ($\Delta H$) is a measure of the relaxation rate of spin fluctuation and hence the temperature and angular dependence of $\Delta H$ exhibit anomalies near $T_N$ {\color{blue}\cite{Seehra1972}}. At lower temperatures, the \RK{linewidth} increases down to the lowest measured temperature of $T$ = 1.8 K. Such divergence of the \RK{linewidth} is observed as the $T_N$ is approached due to dominating spin fluctuation near the critical region. Therefore, such critical broadening on approaching the antiferromagnetic transition indicates the importance of the increasing critical fluctuation close to antiferromagnetic phase transition, and is a characteristic feature of a second order phase transition. In case of the SrCu(OH)$_3$Cl, such critical broadening emerges below $T$ $\sim$ 3, 4 and 3 K along $a$, $b$ and $c$ directions respectively, which are around 3$T_N$. The temperature dependent linewidth follows a power law at low temperatures given by $(1-\frac{T}{T_N})^{-p}$, where $T_N$ is the transition temperature, and the exponent $p$ defines the critical behavior {\color{blue}\cite{Heinrich2003}}. In Fig. {\color{blue}\ref{EsrCritical}(a, b, c)}, we show $\Delta H$ as a function of $(1-T/T_N)$ in log-log scale, which is linear. It indicates that $\Delta H$ exhibits power law divergence close to $T_N$. 

It is interesting that the angular dependence of the linewidth is also temperature dependent. In Fig. \ref{EsrAngle} we show the angular variation of the ESR linewidth in the $b$-$c$ plane at $T$ = 1.8 and 4.2~K. Note that the angular variation of the linewidth at $\theta$=0$^\circ$ (which corresponds to the applied field is parallel to the crystallographic $b$ axis) is sharper at lower temperature. At $T$ = 4.2~K, the angle dependence of the linewidth reveals a $(cos^2\theta+1)$ dependence of the angle, $\theta$ is the angle between the  direction of the static magnetic field and the crystalline axes. As the temperature approaches $T_N$, this term alone is not sufficient to describe the angular variation of the linewidth. At $T$ = 1.8~K, an additional term given by $(3cos^2\theta-1)^2$ {\color{blue}\cite{Riedel1975}} gives satisfactory fit as shown in Fig. \ref{EsrAngle}. The angular dependence of the linewidth showing $(cos^2\theta+1)$ behavior is regularly encountered in weakly correlated or three-dimensional exchange narrowed spin systems. The additional contribution at $T$= 1.8~K may be associated with an additional broadening due to critical fluctuation close to the antiferromagnetic transition. In the case of 2-D spin system, such an angular variation of the linewidth has been observed, which is caused by the increasing dominance of the long-wavelength fluctuations in low dimensional magnets {\color{blue}\cite{Richards1974,Chehab1991}}. Therefore, it may be a possible indication of a dimensional crossover from zero dimensional system at high temperatures to a higher dimensional spin system at low temperatures as SrCu(OH)$_3$Cl undergoes a phase transition below $T_N$ from an isolated triangular spin system.

\subsubsection{\label{sec:level}$g$-factor anomaly}
As mentioned before, the unusual asymmetric shape of the ESR spectra in X-band frequency gives rise to large uncertainties in the ESR parameters, including the $g$-factor. The magnitude and the temperature dependence of the distortion parameter $\sigma$ are quite distinct along the three directions, as shown in the inset of Fig. \ref{EsrIntens}\textcolor{blue}{(c)}. In the investigated temperature range, the magnitude of $\sigma$ is large along $b$- and $c$-directions compared to the $a$-axis. Interestingly, in the temperature variation of $\sigma$ we can distinguish three temperature regions, where $\sigma$ shows different temperature dependence. At low temperatures, $\sigma$ changes rapidly with increase in the temperatures until $T$ = 3-5 K, which is followed by a plateau region in between 5-10 K, followed by rapid change again on further increase in temperature. Such behaviour is prominently observed particularly along $H_{dc}$ $||$ $b$ and $c$ axis, whereas it is less pronounced for $H_{dc}$ $||$ $a$.

Nevertheless, in the main panel of Fig. \ref{EsrIntens}\textcolor{blue}{(c)}, we have shown the temperature variation of the apparent g-factor obtained from resonance field ($H_0$) and microwave frequency ($\nu$) via the Larmor condition $h\nu = g\mu_B H_{0}$, where $\nu$ = 9.47 GHz is the microwave frequency, and $H_0$ is obtained from the fitting of the ESR spectra using Eq. (\ref{ESReq}). As shown in Fig. \ref{EsrSpectra}\textcolor{blue}, the ESR resonance field along $H_{dc}$ $||$ $b$-axis is much larger than a and c-axis at low temperatures. The apparent g-factor along all three directions also exhibits anomalous temperature dependence. At high  temperatures, g-factor is nearly insensitive to the temperature, however, at low temperatures it changes rapidly as the $T_N$ is approached. It is important to note that  the uncertainty of the g-value becomes larger as the linewidth becomes comparable to the resonance field of the ESR spectra. Therefore, one has to consider the strong increase of the linewidth with temperatures.  For that reason, we assume the uncertainty in the resonance field as 5\% of the line width and obtain the error bars shown in Fig. \ref{EsrIntens}\textcolor{blue}{(c)} {\color{blue}\cite{Zakharov2006}}. To further understand the spin-dynamics detail ESR measurements at higher frequencies are required.
\section{\label{sec:level}Summary}
We have studied the magnetic and thermodynamic properties of the single crystalline samples of the quantum spin trimer compound SrCu(OH)$_3$Cl using ac and dc magnetic susceptibility, specific heat, NMR and X-band ESR measurements. While the dc magnetic measurement does not show any phase transition down to  2 K, the specific heat measurement exhibit $\lambda$-like sharp anomaly at $T_{\rm N}$ = 1.2~K at zero magnetic field. The anomaly is progressively suppressed at higher fields. This indicates that the present compound undergoes an antiferromagnetic transition at low temperature. At higher fields, additional features appear in the specific heat, and we have \RK{constructed} a $H$-$T$ phase diagram based on the effect of magnetic field on the specific heat. We can describe the temperature dependent magnetization, the field dependant magnetization and the high field specific heat uniquely well with an ideal equilateral isolated triangle simulation, \RK{despite the observation that two different Cu atoms are present in the triangles}. We have further characterized the ground state and the phase transition by the NMR and electron spin resonance measurements. The $T_1^{-1}$ in $^1$H NMR spectra exhibits sharp reduction below $T$ $\sim$ 15 K, which indicates towards either dominating magnetic fluctuations or any structural anomaly. The ESR intensity does not reveal any signature of nearby phase transition, however, the ESR line width rapidly increases at low temperatures which indicates the important role of the critical spin fluctuations above $T_N$. Overall this study proves SrCu(OH)$_3$Cl to be an ideal model system of isolated isosceles triangles in the range of 2-300 K, well captured by theoretical modeling. This allows less directly interpreted techniques such as thermal transport to test their physical understanding in the near future.
\section{\label{sec:level} Acknowledgment}
This work was supported by the Deutsche Forschungsgemeinschaft (DFG). We acknowledge support by the DR228/68-1. We also \RK{are indebted to} Mrs. G. Unteiner and Mrs. E. Br\"ucher for their technical assistance and M. Isobe for the use of the Rigaku XRD. Work of P.D. at TU Wien was funded by the Czech Science Foundation (research project GA\v{C}R 23-06810O). \RK{S.B. thanks the CNRS research infrastructure INFRANALYTICS (FR 2054) for support}.

\bibliography{bib}
\end{document}